\documentclass[10pt,final,letterpaper]{IEEEtran}
\usepackage{amsmath}
\usepackage{latexsym}
\usepackage{graphicx}
\usepackage{bbding}
\usepackage{indentfirst}
\usepackage{cases}
\usepackage{algorithm2e}
\usepackage{subeqnarray}
\IEEEoverridecommandlockouts

\begin{document}
\title{Interference-Aware Resource Control in Multi-Antenna Cognitive Ad Hoc Networks with Heterogeneous Delay Constraints}
\author{\authorblockN{Xiaoming~Chen,~\IEEEmembership{Member,~IEEE,} and Hsiao-Hwa~Chen,~\IEEEmembership{Fellow,~IEEE}
\thanks{This work was supported by the grants from the NUAA Research
Funding (NN2012004), the open research fund of State Key
Laboratory of Integrated Services Networks, Xidian University (No.
ISN14-01), and Taiwan National Science Council
(NSC99-2221-E-006-016-MY3).}
\thanks{Xiaoming~Chen (e-mail: {\tt chenxiaoming@nuaa.edu.cn}) is
with the College of Electronic and Information Engineering,
Nanjing University of Aeronautics and Astronautics, and also with
the State Key Laboratory of Integrated Services Networks, Xidian
University, China. Hsiao-Hwa~Chen (e-mail: {\tt
hshwchen@ieee.org}) is with the Department of Engineering Science,
National Cheng Kung University, Tainan, 70101 Taiwan.}}}

\markboth{IEEE Communications Letters, Vol. XX, No. Y, Month 2013}{}
\maketitle

\begin{abstract}
In this work, we consider a multi-antenna cognitive ad hoc network
(CAHNet) with heterogeneous delay requirements. To fulfill the
interference and delay constraints simultaneously, we propose to
perform adaptive zero-forcing beamforming (ZFBF) at cognitive
transmitters according to interference channel state information
(CSI). To assist the CAHNet to obtain the interference CSI, we use
a win-win inter-network cooperation strategy, namely quantized
interference CSI feedback from the primary network to CAHNet
through a feedback link, under the condition that the CAHNet pays
a proper price for it. Considering the scarcity of feedback and
power resources, we focus on the minimization of the overall
resource cost subject to both interference and delay constraints.
To solve the problem, we derive a joint feedback and power control
algorithm amongst multiple links of CAHNet. Finally, simulation
results validate the effectiveness of the proposed algorithm.
\end{abstract}

\IEEEpeerreviewmaketitle

\section{Introduction}
Wireless communication with a given delay guarantee is a challenging
issue, especially in cognitive networks. On one hand, more
resources, such as power and spectrum, should be allocated to
satisfy the delay requirement over a fading channel. On the
other hand, the utilization of the resource is strictly limited
due to spectrum scarcity and interference constraint in a
cognitive network. Therefore, resource allocation in a cognitive
network with a given delay guarantee receives a lot of attention.

Recently, cognitive ad hoc network (CAHNet) has become an active
research topic due to its ability of self-adaptation to wireless
channel conditions, self-exploration of available spectrum, and
self-organization of ubiquitous interconnection \cite{Cogad1}
\cite{Cogad2}. Resource control in a CAHNet with a delay
constraint poses a great challenge different from those in a
normal cognitive network. Let us take a look at a simple example.
Since aggregated interference to a primary network is determined
by multiple links of CAHNet, power control works based not only on
the interference constraint, but also on the interaction of
different links. Specifically, if one link is allowed to produce
more interference, it can use high power to meet its delay
requirement, while the other links must lower the power due to a
small interference proportion for a given total interference
constraint. In other words, the power control should take into
account heterogeneous delay requirements. However, it seems to be
difficult to satisfy the delay requirements by power control alone
in a CAHNet, because transmit power is confined by an interference
upper bound. Inspired by interference cancellation in
multi-antenna systems, it is found that the performance of a
multi-antenna cognitive network can be further improved by
exploiting the spatial degrees of freedom. If a cognitive network
has perfect interference CSI, the interference can be cancelled
completely or confined within a required range by using a proper
transmit beam \cite{MultiantennaCog1}. In \cite{MultiantennaCog2},
the authors presented an optimal beam design method to maximize
the spectrum efficiency, while satisfying the interference
constraint. If a cognitive transmitter has only partial
interference CSI, the authors in \cite{MultiantennaCog3} and
\cite{LimitedFeedback} proved that it is also beneficial to
improve the performance through robust beamforming. However, the
previous works did not solve the problems on how a cognitive
network obtains the interference CSI and what is the relationship
between the performance and the amount of interference CSI in a
cognitive network.

In this letter, we propose to perform adaptive beamforming at
multiple cognitive transmitters based on the interference CSI to
satisfy both interference and delay requirements. Especially, we
suggest to use limited cooperation between two networks to achieve
a win-win situation accordingly. Specifically, a CAHNet purchases
some quantized interference CSI from a primary network at a proper
price, so that the CAHNet fulfills the delay requirement and the
primary network gets some rewards. Furthermore, we reveal the
relationship between the inter-network interference, delay
requirement, feedback bits, and transmit power. Then, via
minimizing the overall resource cost while satisfying interference
and delay requirements, we derive a joint feedback and power
control algorithm, which can achieve the same effect as an optimal
exhaustive search algorithm, but at a relatively low complexity.

The rest of this letter is outlined as follows. Section II gives a
brief introduction of a multi-antenna CAHNet with heterogeneous
delay requirements. Section III focuses on the design of a joint
feedback and power control algorithm, and Section IV presents the
numerical results to validate the performance of the proposed
algorithm. Finally, Section V summarizes the findings obtained in
this letter.

\begin{figure}[h] \centering
\includegraphics [width=0.4\textwidth] {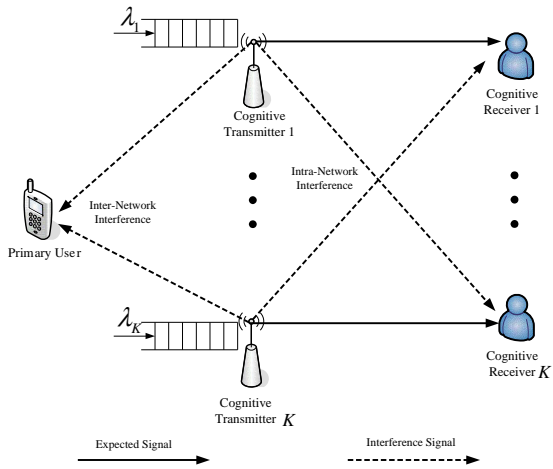}
\caption {An overview of a cognitive ad hoc network model.}
\label{Fig1}
\end{figure}

\section{System Model}

Let us consider a cognitive ad hoc network (CAHNet) consisting of
$K$ transmitter-receiver links coexisting with a primary network,
as depicted in Fig. \ref{Fig1}. For ease of analysis, we assume
that each cognitive transmitter equips with $N_t$ antennas and
each cognitive receiver has one antenna. We also assume that the
primary base station is far from the CAHNet, so we can omit the
primary base station here. A primary user deploys one antenna due
to the size limitation. We use $\textbf{h}_{j,i}$ to denote the
channel from cognitive transmitter $i$ to cognitive receiver $j$,
and $\textbf{g}_{0,i}$ to denote the channel from cognitive
transmitter $i$ to the primary user. Assume that all the channels
remain constant during a transmission interval and suffer
independent fading on an interval by interval basis according to a
circular symmetry complex gaussian distribution with zero mean and
variance $\sigma^2_{j,i}$, where $i=1,\cdots,K$ and
$j=0,\cdots,K$. At the beginning of each transmission interval,
cognitive transmitters obtain partial interference CSI from the
primary user by using a codebook-based limited feedback scheme
\cite{LimitedFeedback}. Specifically, the primary user obtains the
instantaneous interference CSI via channel estimation, and selects
an optimal quantized codeword for the interference channel
$\textbf{g}_{0,i}$ from a predetermined quantization codebook
$\mathcal{G}_i=\{\hat{\textbf{g}}_{i,1},\cdots,
\hat{\textbf{g}}_{i,2^{B_i}}\}$ of size $2^{B_i}$ with a criterion
such that
$l=\arg\max_{\hat{\textbf{g}}_{i,l}\in\mathcal{G}_i}|\tilde{\textbf{g}}_{0,i}^H\hat{\textbf{g}}_{i,l}|^2$,
where
$\tilde{\textbf{g}}_{0,i}=\frac{\textbf{g}_{0,i}}{\|\textbf{g}_{0,i}\|}$
is the corresponding channel direction. Then, the primary user
conveys index $l$ to cognitive transmitter $i$ with $B_i$ bits via
a feedback link, and cognitive transmitter $i$ recovers the
quantized CSI from the same codebook. In a CAHNet, we assume that
cognitive transmitter $i$ may interfere at most $L$ closest
cognitive receivers, and it can obtain full directional
information of these intra-network interference channels
$\tilde{\textbf{h}}_{i_j,i}$, where $j=1,\cdots,L$, through an
intra-network information exchange mechanism. For example, the
cognitive receiver estimates the intra-network interference CSI
and conveys it to a central control node, which delivers the CSI
to the corresponding cognitive transmitter. Based on the
information of inter-network and intra-network interference
channels, cognitive transmitter $i$ determines its optimal
transmit beam $\textbf{w}_i$ making use of zero-force beamforming
(ZFBF) method, such that we have
\begin{equation}
\hat{\textbf{g}}_{i,\textrm{opt}}^{H}\textbf{w}_i=0,\label{eqn2}
\end{equation}
\begin{equation}
\textbf{h}_{i_j,i}^H\textbf{w}_i=0,\label{eqn3}
\end{equation}
where $\hat{\textbf{g}}_{i,\textrm{opt}}$ is the optimal quantized
inter-network interference CSI related to cognitive transmitter $i$
for a given transmission interval.

In such a CAHNet, all cognitive links require different transmission
services. The data for link $i$ arrives in the form of packet of
a fixed length $N_b$ bits with its average arrival rate $\lambda_i$
(packets/interval), and it has the minimum average delay requirement
$D_i$ (intervals) related to its service style. Following
\cite{Minimumrate}, in order to satisfy the delay constraint we know that the
average transmission rate should meet the following condition:
\begin{equation}
\bar{R}_i\geq\frac{(2D_i\lambda_i+2)+\sqrt{(2D_i\lambda_i+2)^2-8D_i\lambda_i}}{4D_i}\frac
{N_b}{T}=\bar{r}_i,\label{eqn4}
\end{equation}
where $T$ is the length of a transmission interval.

The precondition that a CAHNet is allowed to access the licensed
spectrum is that the aggregated inter-network interference must
satisfy a given constraint. It was proved that an average
interference constraint (AIC) is more beneficial to improve the
performance of a cognitive network than a peak interference
constraint (PIC) \cite{AIC1}. Thus, similar to \cite{AIC2}, we
take AIC as the interference constraint. In order to realize delay
guaranteed transmission at a minimum cost, this work focuses on
the minimization of the overall resource cost subject to both
interference and delay constraints by jointly optimizing feedback
bits $B_i$ and transmit power $P_i$ for $i=1,\cdots,K$ according
to the interference CSI, delay requirements, and arrival rates.

\section{Delay Driven Joint Resource Control}

As interference constraint is the precondition, prior to discussing
joint resource control, we first perform an investigation on the
relationship of the average aggregated interference, transmit
power, and feedback bits.

According to the theory of random vector quantization \cite{RVQ},
the relationship between the original and quantized interference
channel direction vectors can be expressed as
\begin{equation}
\tilde{\textbf{g}}_{0,i}=\sqrt{1-a_{0,i}}\hat{\textbf{g}}_{i,\textrm{opt}}+\sqrt{a_{0,i}}\textbf{s}_{0,i},\label{eqn5}
\end{equation}
where $\textbf{s}_{0,i}$ is the quantization error due to finite
quantization bits, which is an unit norm vector isotropically
distributed in the null-space of $\hat{\textbf{g}}_{i,\textrm{opt}}$,
and is independent of $a_{0,i}$.
$a_{0,i}=\sin^2(\angle(\tilde{\textbf{g}}_{0,i},\hat{\textbf{g}}_{i,\textrm{opt}}))$
is the magnitude of the quantization error. Therefore, the
inter-network interference caused by cognitive transmitter $i$ is
given by
\begin{equation}
|\sqrt{P_i}\textbf{g}_{0,i}^H\textbf{w}_i|^2=P_i\|\textbf{g}_{0,i}\|^2a_{0,i}|\textbf{s}_{0,i}^H\textbf{w}_i|^2,\label{eqn6}
\end{equation}
which holds because of (\ref{eqn2}). The sum
average interference can be expressed as
\begin{eqnarray}
\bar{I}
&=&\sum\limits_{i=1}^{K}P_iE[\|\textbf{g}_{0,i}\|^2]E[a_{0,i}]E[|\textbf{s}_{0,i}^H\textbf{w}_i|^2],\label{eqn7}
\end{eqnarray}
which follows due to the fact that the magnitude and
direction of a channel are independent of each other. Examining
the right hand side of (\ref{eqn7}), we can find that
$\|\textbf{g}_{0,i}\|^2$ is $\sigma_{0,i}^2\chi^2(2N_t)$
distributed, so that we have
$E[\|\textbf{g}_{0,i}\|^2]=N_t\sigma_{0,i}^2$. According to the
definition of quantization error, $a_{0,i}$ is equal to
$1-|\tilde{\textbf{g}}_{0,i}^H\hat{\textbf{g}}_{i,\textrm{opt}}|^2$.
For an arbitrary quantization codeword $\hat{\textbf{g}}_{i,l}$,
$1-|\tilde{\textbf{g}}_{0,i}^H\hat{\textbf{g}}_{i,l}|^2$ is
$\beta(N_t-1,1)$ distributed, so that $a_{0,i}$ is the minimum of
$2^{B_i}$ independent $\beta(N_t-1,1)$ random variables, and thus
its expectation can be tightly upper bounded as
$E[a_{0,i}]<2^{-\frac{B_{i}}{N_t-1}}$. Finally, since
$\textbf{s}_{0,i}$ and $\textbf{w}_{i}$ are i.i.d. isotropic
vectors in a $(N_t-1)$ dimensional null-space of
$\hat{\textbf{g}}_{i,\textrm{opt}}$,
$|\textbf{s}_{0,i}^H\textbf{w}_{i}|^2$ is $\beta(1, N_t-2)$
distributed, whose expectation is equal to $\frac{1}{N_t-1}$. As a
result, we have $\bar{I}<\sum_{i=1}^K\frac{\sigma^2_{0,i}
P_iN_t}{N_t-1}2^{-\frac{B_i}{N_t-1}}$. In other words, for a given
AIC $\bar{I}_{\textrm{AIC}}$, the power and feedback joint control
must satisfy $\sum_{i=1}^K\frac{\sigma^2_{0,i}
P_iN_t}{N_t-1}2^{-\frac{B_i}{N_t-1}}\leq\bar{I}_{\textrm{AIC}}$.

For cognitive link $i$, the receive signal can be expressed as
\begin{eqnarray}
y_i&=&\sqrt{P_i}\textbf{h}_{i,i}^H\textbf{w}_ix_i+\sum\limits_{j=1}^L\textbf{h}_{i,i_j}^H\textbf{w}_{i_j}x_{i_j}+n_i\nonumber\\
&=&\sqrt{P_i}\textbf{h}_{i,i}^H\textbf{w}_ix_i+n_i,\label{eqn8}
\end{eqnarray}
where $x_i$ is the normalized transmit signal of cognitive
transmitter $i$, $n_i$ is additive white Gaussian noise with zero
mean and unit variance. (\ref{eqn8}) follows from (\ref{eqn3}).
Hence, the receive SNR and the corresponding transmission rate are
$\gamma_i=P_i|\textbf{h}_{i,i}^H\textbf{w}_i|^2$ and
$R_i=\log_2(1+\gamma_i)$, respectively. Since $\textbf{w}_i$ is a
normalized vector independent of $\textbf{h}_{i,i}$, $\gamma_i$
is $P_i\sigma_{i,i}^2\chi^2(2)$ distributed. In this context, the
average transmission rate for cognitive link $i$ can be computed as
\begin{eqnarray}
\bar{R}_i&=&W\int_0^{\infty}\log_2(1+x)\frac{\exp\left(-\frac{x}{P_i\sigma_{i,i}^2}\right)}{P_i\sigma_{i,i}^2}dx\nonumber\\
&=&W\log_2(e)\exp\Big(\frac{1}{P_i\sigma_{i,i}^2}\Big)E_1\Big(\frac{1}{P_i\sigma_{i,i}^2}\Big),\label{eqn9}
\end{eqnarray}
where $W$ is the bandwidth of licensed spectrum, and $E_1(x)$ is
the exponential-integral function of the first order. For a given
arrival rate $\lambda_i$, in order to fulfill delay constraint
$D_i$, average transmission rate $\bar{R}_i$ should not be less
than $\bar{r}_i$, as shown in (\ref{eqn4}). Since $\bar{R}_i$ is an
increasing function of $P_i$, the delay constraint is equivalent
to $P_i\geq P_i^{\star}$, where $P_i^{\star}$ is the solution of
the function
$W\log_2(e)\exp(\frac{1}{x\sigma_{i,i}^2})E_1(\frac{1}{x\sigma_{i,i}^2})=\bar{r}_i$.

In a CAHNet, transmit power and feedback bits are two important
and scarce resources, where transmit power affects terminal battery life
time directly and feedback bits determine the cost. Thus, this
letter focuses on the design of a joint power and feedback control
algorithm by minimizing the overall resource cost while satisfying
interference and delay constraint. Let $\mu$ be the feedback price
per bit from a primary user to cognitive transmitters, which
depends on the feedback cost consisting of the power and spectrum
used for channel estimation, codeword selection and index
conveyance, and $\varphi$ be the power utilization cost per watt
at cognitive transmitters. Then, the joint resource control is
equivalent to the following optimization problem:
\begin{eqnarray}
J_1:&& \min\limits_{\textbf{B},\textbf{P}}\sum\limits_{i=1}^{K}\mu
B_i+\varphi P_i,\label{eqn10}\\
s.t.&&\sum\limits_{i=1}^K\frac{\sigma^2_{0,i}
P_iN_t}{N_t-1}2^{-\frac{B_i}{N_t-1}}\leq\bar{I}_{\textrm{AIC}},\label{eqn11}\\
&&P_i\geq P_i^{\star},~~i=1,\cdots,K, \label{eqn12}
\end{eqnarray}
where $\textbf{B}=\{B_1,\cdots,B_K\}^T$ and
$\textbf{P}=\{P_1,\cdots,P_K\}^T$ are $K$ dimensional feedback
bits and transmit power vectors, respectively. (\ref{eqn10}) is
the overall resource cost function, and (\ref{eqn11}) and
(\ref{eqn12}) represent interference and delay constraints,
respectively. Clearly, $J_1$ is a mixed integer programming
problem, and it is difficult to get a close-form expression of
optimal $\textbf{B}$ and $\textbf{P}$. Examining the constraint
condition (\ref{eqn11}), we can find that the lower the power, the
smaller the interference. Thus, we can use less bits to meet AIC.
In other words, the resource cost is minimized when using the minimum
power. As shown in (\ref{eqn12}), $P_i^{\star}$, where $i=1,\cdots,K$, are
the lower bound on the power due to the delay constraint.
Therefore, they give the optimal transmit power indeed. Under this
condition, $J_1$ is reduced to
\begin{eqnarray}
J_2:&&
\min\limits_{\textbf{B}}\sum\limits_{i=1}^{K} B_i,\label{eqn14}\\
s.t.&&\sum\limits_{i=1}^K\frac{\sigma^2_{0,i}
P_i^{\star}N_t}{N_t-1}2^{-\frac{B_i}{N_t-1}}\leq\bar{I}_{\textrm{AIC}},\label{eqn15}
\end{eqnarray}
where $J_2$ is a pure feedback control problem. Intuitively,
allocating more bits to a cognitive transmitter with a strong
interference mitigation capability is beneficial to decrease the
feedback cost. Inspired by this idea, we propose a greedy feedback
control algorithm. First, we define feedback gain as the amount of
interference reduction by adding one more bit feedback. For example,
the feedback gain of cognitive transmitter $i$ with $B_i$ feedback
bits at present is $\eta_i(B_i)=\frac{\sigma^2_{0,i}
P_i^{\star}N_t}{N_t-1}(2^{-\frac{B_i}{N_t-1}}-2^{-\frac{B_i+1}{N_t-1}})$.
Note that feedback gain is variable, as inter-network interference
is not a linear function of feedback bits. Then, we can compare
the feedback gains of all transmitters and add one more bit to the
transmitter with the largest feedback gain. Repeat the process
until the aggregated interference satisfies the AIC. The joint
feedback and power control algorithm can be summarized as follows:
\begin{enumerate}
\item Initialization: With given $N_t$, $K$, $\bar{I}_{\textrm{AIC}}$, $W$, $T$, $D_i$,
$N_b$ and $\sigma_{j,i}^2$, set $\textbf{P}=\textbf{0}$ and
$\textbf{B}=\textbf{0}$.
\item
Compute $\bar{r}_i$, and then $P_i^{\star}$. $P_i=P_i^{\star}$ for
$i=1,\cdots,K$ are the optimal transmit powers.
\item
Let $\eta_i(B_i)=\frac{\sigma^2_{0,i}
P_i^{\star}N_t}{N_t-1}(2^{-\frac{B_i}{N_t-1}}-2^{-\frac{B_i+1}{N_t-1}})$.
Search $i^{\star}=\arg\max_{1\leq i\leq K}\eta_i(B_i)$. Let
$B_{i^{\star}}=B_{i^{\star}}+1$.
\item
If $\sum_{i=1}^K\frac{\sigma^2_{0,i}
P_i^{\star}N_t}{N_t-1}2^{-\frac{B_i}{N_t-1}}>\bar{I}_{\textrm{AIC}}$,
then go to 3).
\end{enumerate}

Assume that the total feedback amount is $B$ bits. The
computation complexity of the above greedy resource control
algorithm scales with $KB$, while that of the optimal algorithm
based on exhaustive search scales with $K^B$, so that the proposed
algorithm has a much lower complexity, especially with a larger
$K$. In some extreme cases shown as follows, it is likely to
obtain the optimal control algorithms more easily.
\begin{enumerate}
\item Loose AIC and/or long separation distance from a primary
user: If AIC is sufficiently loose, so that the AIC can always be
satisfied even when the design of transmit beam at a cognitive
transmitter is independent of inter-network interference CSI. In
other words, without interference CSI, a CAHNet can work well.
Under this condition, $P_i=P_i^{\star}$ and $B_i=0$ for all $i$
are the optimal control algorithm. Furthermore, in the case of
$P_i=P_i^{\star}$ and $B_i=0$, the aggregated average interference
is $\bar{I}_0=\sum_{i=1}^K\frac{\sigma^2_{0,i}
P_i^{\star}N_t}{N_t-1}$, such that when
$\bar{I}_{\textrm{AIC}}\geq\bar{I}_0$, the above algorithm is
optimal. Moreover, if the distance between CAHNet and a primary
user is long enough, due to path loss, the aggregated average
interference always satisfies AIC even with $B_i=0$. Let
$\sigma^2_{0,1}=\cdots=\sigma^2_{0,K}=\sigma^2_0=d^{-\alpha}$,
where $d$ denotes the distance and $\alpha$ denotes the path loss
exponent. When
$\sigma^2_0\leq\bar{I}_{\textrm{AIC}}/(\frac{N_t}{N_t-1}\sum_{i=1}^{K}P_{i}^{\star})$,
the AIC is always met. Hence, when the minimum distance between
cognitive transmitters and a primary user is not shorter than
$d=(\bar{I}_{\textrm{AIC}}/(\frac{N_t}{N_t-1}\sum_{i=1}^{K}P_{i}^{\star}))^{1/\alpha}$,
$P_i=P_i^{\star}$ and $B_i=0$ for all $i$ are the optimal control
algorithm. \item $\bar{I}_{\textrm{AIC}}=0$: If a primary user can
not bear any interference from the CAHNet, namely
$\bar{I}_{\textrm{AIC}}=0$, then a cognitive transmitter must know
full interference CSI, and transmits its signal in the null-space
of interference channel. In this case, $B_i$ must approach to
infinity. However, it is impractical in any real CAHNet.
\end{enumerate}

\section{Numerical Results}

To evaluate the performance of the proposed algorithm, we present
several numerical results with various AICs. We set
$N_t=4$, $K=3$, $N_b=80$, $W=50$KHz, $\lambda_1=0.3$,
$\lambda_2=0.4$, $\lambda_3=0.5$, $D_1=2$, $D_2=4$, $D_3=20$,
$\sigma_{1,1}^2=\sigma_{2,2}^2=\sigma_{3,3}^2=0.01$,
$\sigma_{0,1}^2=0.0001$, $\sigma_{0,2}^2=0.0005$,
$\sigma_{0,3}^2=0.0010$, $\mu_1=\mu_2=\mu_3=1$, and $T=5$ ms. In what
follows, we use GA and EA to denote the proposed greedy algorithm
and exhaustive search algorithm, respectively.

\begin{table}\centering
\caption{Optimal number of feedback bits with different AICs.} \label{Tab1}
\scriptsize
\begin{tabular}{|c|c|c|c|c|}\hline
& & $B_1$ & $B_2$ & $B_3$  \\
\hline
& $\textmd{GA}$ & 1 & 6 & 9 \\
\cline{2-5}$\bar{I}_{\textrm{AIC}}=0.01$   & $\textmd{EA}$ & 0 & 6 & 10 \\
\hline
& $\textmd{GA}$ & 0 & 3 & 5 \\
\cline{2-5}$\bar{I}_{\textrm{AIC}}=0.02$   & $\textmd{EA}$ & 0 & 2 & 6\\
\hline
& $\textmd{GA}$ & 0 & 1 & 3\\
\cline{2-5}$\bar{I}_{\textrm{AIC}}=0.03$   & $\textmd{EA}$ & 0 & 0 & 4\\
\hline
& $\textmd{GA}$ & 0 & 0 & 1\\
\cline{2-5}$\bar{I}_{\textrm{AIC}}=0.04$   & $\textmd{EA}$ & 0 & 0 & 1\\
\hline
\end{tabular}
\end{table}

\begin{figure}[h]
\centering \hspace{-0.25in}
\includegraphics [width=0.45\textwidth] {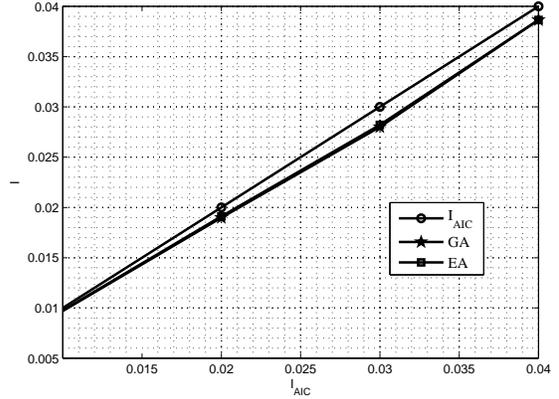}
\caption {Average interference of GA and EA schemes.} \label{Fig2}
\end{figure}

Table {\ref{Tab1}} shows the feedback control results based on GA and
EA with different AICs. As AIC becomes relatively loose, the required feedback
bits decrease sharply. For example, there are 15 bits reduction when
$\bar{I}_{\textrm{AIC}}$ changes from $0.01$ to $0.04$. For each
AIC, cognitive transmitter $3$ is always allocated the most bits,
since its feedback gain is the largest in most cases, which is
consistent with the principle of the proposed greedy control
algorithm. It is found that, although the control results are
different with a given AIC, the total feedback bits are kept the same. In
other words, the optimal control result is not unique, and GA is
also optimal in the sense of resource cost. Moreover, Fig. \ref{Fig2}
presents the average sum interference of GA and EA schemes
under different AICs. It is found that GA has nearly the same
average sum interference as EA. However, it has a quite lower
complexity compared to the exhaustive search algorithm, which is
more appealing in a practical CAHNet.

\section{Conclusion}
A major contribution of this letter is the introduction of the
limited inter-network cooperation into a multi-antenna CAHNet to
achieve a win-win situation. Considering the scarcity of feedback
and power resources, we proposed a greedy resource control
algorithm by minimizing the overall resource cost. The
numerical results confirmed the effectiveness of the proposed
algorithm.

\vfill
\end{document}